\documentclass[apj,graphicx]{emulateapj}
\usepackage{apjfonts}

\newcommand\sax{SAX~J1808.4$-$3658}
\newcommand\suz{{\it Suzaku}}
\newcommand\xmm{{\it XMM-Newton}}

\begin{document}

\title{ 
Broad relativistic iron emission line observed in SAX~J1808.4$-$3658}
\shortauthors{Cackett et al.}
\shorttitle{Broad Iron Line in \sax}

\author{E.~M.~Cackett\altaffilmark{1,2}, 
D.~Altamirano\altaffilmark{3},
A.~Patruno\altaffilmark{3},
J.~M.~Miller\altaffilmark{1},
M.~Reynolds\altaffilmark{1},
M.~Linares\altaffilmark{3},
R.~Wijnands\altaffilmark{3}
}
\email{ecackett@umich.edu}
\altaffiltext{1}{Department of Astronomy, University of Michigan, 500 Church St, Ann Arbor, MI 48109-1042, USA}
\altaffiltext{2}{Chandra Fellow}
\altaffiltext{3}{Astronomical Institute `Anton Pannekoek', University of Amsterdam, Kruislaan 403, 1098 SJ, Amsterdam, the Netherlands}

\begin{abstract}

During the September-October 2008 outburst of the accreting millisecond pulsar \sax{}, the source was observed by both \suz{} and \xmm{} approximately 1 day apart.  Spectral analysis reveals a broad relativistic Fe K$\alpha$ emission line which is present in both data-sets, as has recently been reported for other neutron star low-mass X-ray binaries.  The properties of the Fe K line observed during each observation are very similar.  From modeling the Fe line, we determine the inner accretion disk radius to be $13.2 \pm 2.5$ GM/c$^2$.  The inner disk radius measured from the Fe K line suggests that the accretion disk is not very receded in the hard state. If the inner disk (as measured by the Fe line) is truncated at the magnetospheric radius this implies a magnetic field strength of $\sim3\times10^8$ G at the magnetic poles, consistent with other independent estimates.  
\end{abstract}

\keywords{stars: neutron --- X-rays: binaries --- X-rays: individual: SAX~J1808.4$-$3658}

\section{Introduction}

The accreting millisecond pulsar \sax{} was the first X-ray binary discovered to pulsate in the millisecond range \citep{wijnands98}.  It is a transient compact binary with a $\sim$2.01~hr orbital period \citep{chakra98} and an outburst recurrence time of approximately 2.5 years.  \sax{} has been seen in outburst six times since 1996, the most recent being in September-October 2008.  The presence of pulsations implies the accreting gas is channeled onto the magnetic poles producing hot spots and accretion shocks.  The rotation of the neutron star modulates this emission producing pulsations that add to the unpulsed accretion disk and Comptonized emission \citep[see for example][]{poutanen03}. 

A tool to examine \sax{} is the presence of an iron K-shell fluorescence emission line in its X-ray spectrum, similar to those observed in many other X-ray binaries and AGNs \citep[see][for comprehensive reviews]{miller07,fabian00}. It is generally thought that the shape of iron-K emission line formed in the inner region of accretion disks around black holes (both stellar-mass and in AGN) and neutron stars is sensitive to the distance of the line forming region from the compact object -- as this region gets closer to the compact object relativistic Doppler effects and gravitational redshifts become stronger, producing an asymmetric profile \citep{fabian89}.  The iron K emission line can therefore provide an important way to examine the inner accretion disk.  While this line has long been used to study the inner disk around black holes \citep[e.g.,][]{miller07}, only recently has it been shown that the lines in neutron star low-mass X-ray binaries also display the characteristic asymmetric shape too \citep{bhatta07,cackett08}. The detection of an iron line in an accreting millisecond pulsar is useful for constraining the magnetospheric radius and hence the strength of the magnetic field in this neutron star. 

During the Sept/Oct 2008 outburst, \sax{} was observed by both \suz{} and \xmm{}, approximately 1 day apart.  Initial brief reports of the Fe K line seen by both telescopes were given by \citet{cackett08atel,papittoatel08}.  Further analysis of the \xmm{} data has been presented by \citet{papitto08}.  Here, we present a joint analysis of both \suz{} and \xmm{} data that yields the tightest constraints on the Fe K emission line present in the spectra.

\section{Data Reduction and Analysis}

\begin{figure}
\includegraphics[width=8cm]{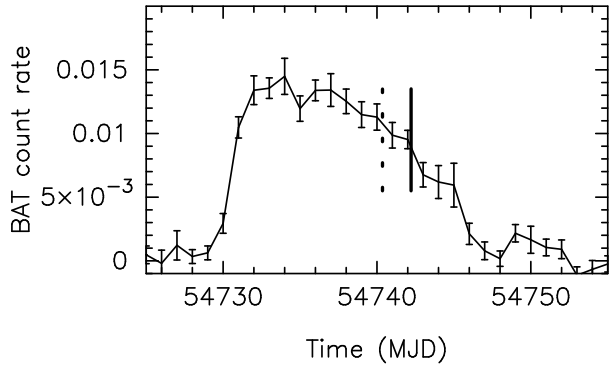}
\caption{{\it Swift}/BAT 15--50 keV light curve of the September-October 2008 outburst of \sax. The dashed and solid lines mark the mid-points of the \xmm{} and \suz{} observations respectively.}
\label{fig:lc}
\end{figure}

The light curve of the Sept/Oct 2008 outburst is given in Fig.~\ref{fig:lc}.  The daily average count rates are taken from the {\it Swift}/BAT Hard X-ray Transient Monitor in the 15-50 keV energy range.  The times when the \suz{} and \xmm{} observations occurred are also indicated.  Below we describe the data reduction for both the \suz{} and \xmm{}, before detailing our spectral analysis.

\subsection{\suz{} data}

\suz{} observed \sax{} starting at 2008 October 2, 16:32 UT, for approximately 42.5 ks (ObsID 903003010).  The XIS was operated in 1/4 window mode, with a 1-sec burst option, and the telescope was pointed at the nominal XIS position. The XIS detectors, which cover the energy range from approximately 0.5-10 keV, were operated in both 3x3 and 5x5 edit modes during the observation.  For each working XIS unit (0, 1 \& 3), we extracted spectra from the cleaned event files using \verb|xselect|.  For the front-illuminated detectors (XIS 0 and 3), the extraction region used was a box of size 270\arcsec{} by 330\arcsec{} centered on the source.  We choose to use a box rather than a circular extraction region as the used region of the detector is a thin 256 by 1024 pixel strip (approximately 270\arcsec{} by 1080\arcsec) when using 1/4 window mode. The background was extracted from a 250\arcsec{} by 230\arcsec{} box situated towards the end of the 1/4 window so as to be free from source photons.  The response files were generated using the \verb|xisresp| script which uses the \verb|xisrmfgen| and \verb|xissimarfgen| tools.  \verb|xissimarfgen| is used with no binning of the response file and simulating 200000 photons.  We then added the spectra (and averaged the responses) from the two front illuminated detectors (XIS 0 and 3) together using the \verb|addascaspec| tool, as recommended by the \suz{} XIS team.  The summed good exposure time of the XIS 0 + 3 spectrum is 42.4 ks (this takes into account the 1-sec burst option which leads to a 50\% livetime fraction).

In this Letter, we choose not use data from the back-illuminated XIS 1 detector.  As the back-illumination increases the effective area at soft energies, this detector is more prone to pile-up.  We found that pile-up was present in the XIS 1 observation (but not significantly in the other detectors).  We tested using an annulus source extraction region that avoids the piled-up central region of the source, but found that when we did this the signal-to-noise ratio was significantly decreased, especially through the Fe K band (6.4--6.97 keV) of interest here.

In addition to the XIS spectra, we also extract the spectrum from the PIN camera which is part of the Hard X-ray Detector (HXD). We extracted the PIN spectrum from the cleaned event file following the standard analysis threads on the \suz{} website. The PIN non-X-ray background was extracted from the observation-specific model provided by the instrument team, and was combined with the standard model for the cosmic X-ray background.

\subsection{\xmm{} data}

\xmm{} observed \sax{} starting on 2008 September 30,  23:53 UT, for a total of $\sim$63 ks (ObsID 0560180601).  Here we present data from the PN detector, which was operated in timing mode during the observation to prevent pile-up. Processing of the ODF failed using \verb|epproc|, therefore we used the PN event file provided by the \xmm{} team.  We checked for background flares, and found that the first $\sim$5 ks were affected by a flare, therefore we excluded that section of the data from the rest of the analysis.  We extracted the spectrum using \verb|xmmselect|.  In timing mode the detector is continously read-out in the Y direction, leading to a bright streak on the detector.  Thus, the source extraction region was chosen to be of width 24 pixels centered on the source and with the length covering the entire readout streak.  The background spectrum was extracted from a strip at the edge of the observed window.  The good exposure time of the source spectrum was 43.4 ks.  The response files were generated using the \verb|arfgen| and \verb|rmfgen| tools.

\subsection{Spectral Fitting}

We fit the spectra using \verb|XSPEC| version 11 \citep{arnaud96}.  First, we fit both the  \suz{} and \xmm{} spectra separately.  For the \suz{} data we fit the XIS spectrum over the 0.7 - 10 keV energy range, and the PIN spectrum over the 14-45 keV range.  For the continuum model, we used an absorbed multicolor disk-blackbody + a single temperature blackbody and an additional power-law \citep[e.g.][]{lin07,cackett08}.  All components are required by the data and statistically improve the fit. For the absorption we used the \verb|phabs| model.   In the \suz{}/XIS spectrum we found that there is a detector feature at approximately 1.85 keV.  We attribute this to a change in the instrument response that is not yet corrected for by the response files \citep[similar features have been seen in other 1/4 window mode observations, e.g.,][]{miller08,cackett08}. We therefore model the feature with a Gaussian absorption line so that it does not effect the continuum fit. We allowed a constant between the \suz/XIS and \suz/PIN spectra to account for any offset in flux calibration, finding $c = 1.37 \pm 0.05$. Such an offset between XIS and PIN flux can occur in 1/4 window mode because small extraction regions must be used. This is a known problem detailed in the `Calibration Issues' on the \suz{} website. 

Initially, we fit the continuum ignoring the iron line region $4-7$ keV.  When examining this region after having fit the continuum, an iron line was clearly present, thus we added the \verb|diskline| \citep{fabian89} model to fit the line and re-fit all the parameters.  The energy of the line center was constrained to be within the Fe K band range, 6.4 to 6.97 keV.  We left the power-law disk emissivity profile, $\beta$, and inner accretion disk radius, $R_{in}$, as free parameters, while the outer accretion disk radius was fixed at a large value (1000 GM/c$^2$).  The inclination, $i$, of the disk was also a free parameter, yet was constrained to lie within 36 to 67 degrees -- the possible range of inclinations determined by optical observations of \sax{} \citep{deloye08}.  The best-fitting parameters (continuum \& line) are given in Tab.~\ref{tab:model}.

\begin{deluxetable}{lcc}
\tablecolumns{3}
\tablewidth{0pc}
\tablecaption{Spectral parameters from separate fitting}
\tablecomments{All uncertainties are at the 90\% confidence level.  The \suz{} spectra are fit over the range 0.7-45 keV, and the \xmm{} spectra are fit from 1.2-11 keV.}
\tablehead{Parameter & Suzaku & XMM}
\startdata
$N_H$ ($10^{21}$ cm$^{-2}$) & $ 0.46\pm0.06$ & $2.3\pm0.1$ \\
Disk $T_{in}$ (keV) & $ 0.48 \pm 0.01$ & $0.23 \pm 0.01$ \\
Disk normalization & $590 \pm 51$ & $20400^{+21800}_{-8000}$ \\ 
Blackbody, $kT$ (keV) & $1.00\pm0.04$ & $0.40 \pm0.01$ \\
Blackbody normalization ($10^{-3}$) & $1.27\pm0.02$  & $1.25\pm0.06$ \\
Power-law index & $1.93\pm0.02$ & $2.08 \pm 0.01$ \\
Power-law normalization & $0.20\pm0.01$ & $0.33 \pm 0.01 $ \\
$E_{line}$ (keV)  & $6.40^{+0.06}$ &  $6.40^{+0.06}$\\
$\beta$  & $-2.9 \pm 0.3$ &  $-3.0\pm0.2$\\
$i$ $(^{\circ})$ &  $50^{+11}_{-4}$ &  $59\pm4$\\
$R_{in}$ (GM/c$^2$) & $12.7^{+11.3}_{-2.0}$ & $13.0\pm3.8$\\
EW (eV) & $134\pm30$ & $118\pm10$ \\
$\chi^2_\nu (\nu)$ & 1.17 (2550) & 1.09 (1950)
\label{tab:model}
\enddata
\end{deluxetable}

\begin{deluxetable}{lc}
\tablecolumns{2}
\tablewidth{0pc}
\tablecaption{Iron line parameters from joint fitting}
\tablecomments{All uncertainties are at the 90\% confidence level}
\tablehead{Parameter & Suzaku \& XMM}
\startdata
$E_{line}$ (keV)  & $6.40^{+0.03}$ \\
$\beta$  & $-3.05 \pm 0.21$ \\
$i$ $(^{\circ})$ &  $55^{+8}_{-4}$ \\
$R_{in}$ (GM/c$^2$) & $13.2\pm2.5$ \\
EW, Suzaku (eV) & $144^{+14}_{-31}$  \\
EW, XMM (eV) & $113\pm10$ \\
$\chi^2_\nu (\nu)$ & 1.14 (4502) 
\enddata
\label{tab:line}
\end{deluxetable}

\begin{figure}
\includegraphics[width=8cm]{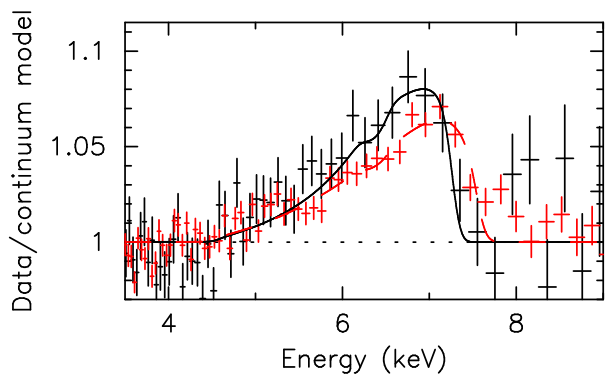}
\caption{Iron K emission line in \sax{} detected by \suz{} (black) and \xmm{} (red).  The ratio of data to continuum model is plotted to show the iron line profile.  The black, solid line shows the best-fitting iron line model to the \suz{} data, and the red, dashed line shows the line model for the \xmm{} data.}
\label{fig:line}
\end{figure}

We used the same method when fitting the \xmm/PN spectrum.  When examining the PN spectrum we found significant residuals below 1.2 keV, therefore, we only fit the spectrum over the 1.2-11 keV energy range.  As with the \suz{} data, a clear Fe K line was present in the data, which we again fitted using the \verb|diskline| model. The best-fitting parameters (continuum \& line) are given in Tab.~\ref{tab:model}, and the Fe K line profiles are shown in Fig.~\ref{fig:line}.

\begin{figure}
\includegraphics[angle=270,width=8cm]{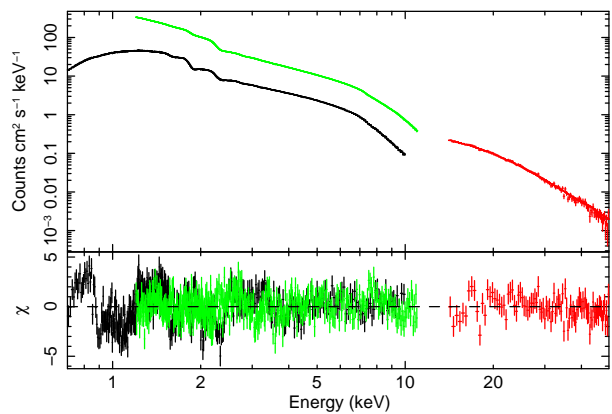}
\caption{\suz{}/XIS 1+3 (black), \suz{}/PIN (red) and \xmm{}/PN (green) spectra of \sax{} (top panel). The residuals of the best fit ($\chi =$ (data-model)/$\sigma$) are plotted in the bottom panel.}
\label{fig:spec}
\end{figure}

The continuum shapes from \suz{} and \xmm{} are different. This is not unexpected given that the observations were non-simultaneous, and the flux (see Fig. \ref{fig:lc}) had changed between the observations.  However, the Fe K line properties are similar (all the parameters are consistent within the uncertainties), thus, we chose to fit the \suz{} and \xmm{} spectra simultaneously to place tighter constraints on the inner disk radius.  In this joint fit, we allowed the continuum parameters to be completely independent, yet tied all the line parameters (except the normalization) between the two datasets.  Parameters from this joint fit are given in Tab.~\ref{tab:line}.  We find an inner accretion disk radius of $R_{in} = 13.2 \pm 2.5$~GM/c$^2$.  The inclination of the disk is measured to be $i = 55^{+8}_{-4}$ degrees.  We show the broadband spectrum in Fig.~\ref{fig:spec}.  

In addition to this phenomenological approach, we also investigated a number of self-consistent spectral fits, including a relativistic disk reflection spectrum.  The \suz{} spectra can be fit well using the \verb|refsch| model instead of a power-law component, leading to a reflection fraction of $0.5\pm0.1$.  However, we note that there is little evidence of the Compton back-scattering hump expected from disk reflection; the HXD/PIN spectrum can be fit well using only a simple power-law.  We will investigate disk reflection in neutron star systems in an upcoming paper.

\subsection{Estimating the magnetic field strength}

If we assume that the inner accretion disk is truncated at the magnetospheric (Alfv\'{e}n) radius (where magnetic pressure balances the ram pressure from infalling material), then from the source luminosity and reasonable assumptions about mass and radius we can estimate the magnetic field strength \citep[for example see Eq 6.19 in][]{fkr}.  Here we modify the formulation of \citet{ibragimov09} substituting $R_{in} = x GM/c^2$ into their Eq. 14 to give the following expression for the magnetic dipole moment:

\begin{eqnarray}
\mu &=& 3.5\times10^{23} \; k_A^{-7/4} \; x^{7/4} \left(\frac{M}{1.4 \; \mathrm{M_\odot}}\right)^2 \nonumber \\ 
&\times& \left( \frac{f_{ang}}{\eta} \frac{F_{bol}}{10^{-9} \; \mathrm{erg \; cm^{-2} \; s^{-1}}} \right)^{1/2}
\frac{D}{3.5 \; \mathrm{kpc}} \;\; \mathrm{G \: cm^3}
\end{eqnarray}

where $\eta$ is the accretion efficiency in the Schwarzschild metric, $f_{ang}$ is the anisotropy correction factor \citep[which is close to unity, see][for details]{ibragimov09}, and the coefficient $k_A$ depends on the conversion from spherical to disk accretion.  Numerical simulations suggesting $k_A = 0.5$ \citep{long05}, and as noted by \citet{ibragimov09}, theoretical models predict $k_A < 1.1$ \citep{psaltis99,kluzniak07}.  For ease of comparison with previous magnetic field estimates, we assume a distance  $D = 3.5\pm0.1$ kpc \citep{galloway06} and mass $M = 1.4 \; \mathrm{M_\odot}$.  

We also use a bolometric conversion factor of $F_{bol}/F_{2-25 keV} = 2.12$ \citep{galloway06}.  The 2-25 keV fluxes that we measure from the \suz{} and \xmm{} observations are $(1.12\pm0.01)\times10^{-9}$ erg cm$^{-2}$ s$^{-1}$ and $(1.16\pm0.01)\times10^{-9}$ erg cm$^{-2}$ s$^{-1}$, respectively.  Here, we just use the average of the two, which leads to $F_{bol} = (2.42 \pm 0.02)\times 10^{-9}$ erg cm$^{-2}$ s$^{-1}$. 

Using $R_{in}$ from the joint \suz/\xmm{} Fe line fit, along with assuming $k_A$ = 1, $f_{ang} = 1$, and $\eta = 0.1$ leads to $\mu = (1.6\pm0.5) \times10^{26}$~G~cm$^{3}$, taking into account the uncertainty in the inner disk radius, flux, and distance. For a stellar radius, $R = 10$ km, this leads to a magnetic field strength of $B = (3.2\pm1.0)\times10^8$~G at the magnetic poles.  Note if we assume $k_A =0.5$, we would infer a larger magnetic field strength approximately a factor of 3 larger.

\section{Discussion}

We have observed a broad, relativistic Fe K emission line in the accreting millisecond X-ray pulsar \sax, using both \suz{} and \xmm{}.  The line is similar in both observations, and by modelling the iron line we determine the inner disk radius, $R_{in} = 13.2 \pm 2.5$ GM/c$^{2}$.  Making reasonable assumptions about the neutron star radius and mass allowed us to estimate the magnetic field strength $B = (3.2\pm1.0)\times10^8$~G at the magnetic poles.  The broadband spectrum from \suz{} displays a hard power-law tail that extends to at least 45 keV.  Although we were able to fit the broadband spectrum well with a reflection model, we note there is no clear sign of the Compton hump that can be seen in black hole systems.  Neutron star LMXB spectra are certainly not as simple as black hole LMXB spectra, with the former often requiring multiple thermal components, which may lead to hide the Compton hump.

In comparison with our inner disk radius, from fitting the \xmm{} data alone, \citet{papitto08} find  $R_{in} = 8.7^{+3.7}_{-2.7}$ GM/c$^{2}$ consistent with our value, though a little smaller.  The small difference may be attributed to the fact that while we fixed the outer disk radius at a large value (so that is not does affect the line profile), \citet{papitto08} left it as a free parameter.  If we also leave the outer disk radius as a free parameter, we reproduce their result.   Of all the parameters of the line model, the line profile is least sensitive to the outer disk radius, $R_{out}$, and this parameter cannot be well constrained from spectral fitting -- for a typical emissivity profile of $R^{-3}$, there is a strong weighting to the innermost part of the disk.  Thus, the slightly different assumptions about the outer disk have only a small effect on the measured inner disk radius.

An independent estimate of the inner accretion disk radius is made by \citet{ibragimov09}.  From {\it RXTE} observations of the 2002 outburst of \sax{}, these authors study the evolution of the pulse profiles, finding a secondary maximum during the late stages of the outburst.  They attribute this to the accretion disk receding and allowing a view of the lower magnetic pole.  Under this assumption, they estimate the inner disk radius, at that specific time in the outburst, to be 19.5 km for $M = 1.4$ M$_\odot$, $R = 12$ km, and $i = 50^\circ$, with larger estimates for larger inclinations.  This is close to our determination of the inner disk radius.

\citet{pandel08} find a broad, relativistic Fe K line in the spectrum of 4U~1636$-$536.  They suggest that the line is a blend of several K$\alpha$ lines in different ionization states.  However, from their modeling of the line profile with 2 ionization state lines (one at 6.4 keV, and one at 7 keV) the inner disk radii they determine for each ionization state are very similar (within 20 GM/c$^2$ of each other).  Yet, it would seem difficult to produce lines with such different ionization states so close to each other.  In addition, the disk emissivity heavily weights the emission from the innermost part of the disk, where the highest ionization state line would come from.  Thus, even if multiple lines are present, it seems reasonable to assume that the highest ionization state line would completely dominate.  Here, we find we do not require a second ionization state to model the line profile.

The power-law index measured during the \suz{} and \xmm{} observations, as well as the power spectrum presented in \citet{papitto08}, is typical of the low-hard state \citep[often referred to as the `island' state, e.g., see][for further details on neutron star LMXB states]{vanderklis06}.  The inner disk radius that we measure from the Fe line is only $\sim$2 times the innermost stable circular orbit (for a Schwarzschild metric), suggesting that there is little recession of the accretion disk in this hard state.  This is in disagreement with the model proposed by \citet{done07} where the disk is recessed in the island state.  A similar result has been seen in black hole LMXBs -- observations of iron lines and/or a cool disk in the low-hard state of black hole LMXBs also suggests that there is little recession of the accretion disk in such a state \citep{miller06b,miller06a}.  Moreover, \citet{rykoff07} show that there is a disk component in the black hole LMXB XTE~J1817$-$330 that cools as $L_x \sim T^4$ as the source transitions to the low-hard state, with an inner disk radius consistent with the innermost stable orbit at all times.

Our estimate of the magnetic field strength is broadly consistent with previous estimates.  From the long-term spin down, \citet{hartman08} determine B $<1.5\times10^8$ G at the magnetic poles.  \citet{disalvo03} combine the quiescent luminosity of \sax{} along with considerations of the magnetospheric radius during quiescence to estimate B $= 1-5\times10^8$ G (at the equator, note that for a dipole the field strength at the equator is a factor of 2 smaller than at the poles).  An earlier estimate by \citet{psaltis99} is less constraining implying a large possible range in surface dipole field of $10^8-10^9$ G at the stellar equator.  Moreover, \citet{ibragimov09} estimate $B = (0.4-1.2)\times10^8$ G from their measurement of the inner disk radius.  The estimation of the magnetic field strength here is, of course, dependent on a number of assumptions. In particular, the coefficient $k_A$ is the most uncertain, with a lower value of $k_A$ increasing our estimate of the magnetic field strength. The uncertainty from our particular choice of $k_A$ is not included in the quoted confidence limit.  Moreover, our estimate is based on the assumption that the disk is truncated at the magnetospheric radius.  This, however, may not necessarily be the case, as MHD simulations have shown that periodicities at the star's rotation frequency can still be present during unstable accretion \citep{kulkarni09}.  Additionally, the presence of the stellar magnetic field affects the inner part of the accretion disk which may in turn affect the iron line profile.  It is not clear how this will change the shape of the line profile from the standard \verb|diskline| model used here. 

Finally, optical observations during the 2008 outburst of \sax{} have lead to an improved mass function for the pulsar, with $M_1\sin^3{i} = 0.48^{+0.17}_{-0.14}$ M$_\odot$ \citep{elebert09}.  From these observations they are unable to constrain the binary inclination any better than previously.  If we assume that the inclination we measure from the Fe K line is the same as the inclination of the binary system, then this would imply quite a low pulsar mass of $0.9^{+0.5}_{-0.4}$ M$_\odot$.  However, there is still a large uncertainty in this mass determination, which is consistent with the canonical 1.4 M$_\odot$.  In contrast, tight constraints on the thermal emission from the neutron star surface during quiescence imply that \sax{} requires enhanced levels of core cooling, which can possibly be achieved through have a larger mass \citep{heinke08}. 

It is also interesting to note that  \citet{elebert09} estimate a shorter distance to \sax{} of $\sim2.5$ kpc based on the equivalent width of the interstellar absorption lines.  This shorter distance is the same as that estimated from type-I X-ray bursts by \citet{intzand01}.  Such a distance would reduce our magnetic field strength estimate by 0.7.

\acknowledgements
EMC gratefully acknowledges support provided by NASA through the Chandra Fellowship Program, grant number PF8-90052.

\end{document}